\documentclass{emulateapj}

\usepackage{latexsym}
\usepackage{mathrsfs}
\usepackage{times} 
\usepackage{graphics}

\usepackage[german,english,american]{babel} 

\newcommand{\xim}{{\tt XIM} }
\newcommand{\ixo}{{\em IXO} }

\begin{document}
\title[High-Resolution X-ray Spectroscopy of Cluster Feedback]{Prospects of
  High-Resolution X-ray Spectroscopy for AGN Feedback in Galaxy
  Clusters} \author{S. Heinz$^1$, M. Br\"{u}ggen$^2$, B. Morsony$^1$}
\address{$^1$Astronomy Department, University of Wisconsin-Madison,
  475. N. Charter St., Madison, WI 53706}
\address{$^2$Jacobs University Bremen, PO Box 750561, 28725 Bremen,
  Germany} \email{heinzs@astro.wisc.edu}

\begin{abstract}
  One of the legacies of the {\rm Chandra} era is the discovery of
  AGN-inflated X-ray cavities in virtually all cool-core clusters,
  with mechanical luminosities comparable to or larger than the
  cluster cooling rate, suggesting that AGN might be responsible for
  heating clusters.  This discovery poses a new set of questions that
  cannot be addressed by X-ray imaging or modeling alone: Are AGN
  actually responsible for halting cooling flows? How is the AGN
  energy transferred to heat? How tight is the observed balance
  between heating and cooling? Using numerical simulations and a new
  virtual X-ray observatory tool, we demonstrate that high-resolution,
  high-throughput X-ray spectroscopy can address these questions and
  that the International X-ray Observatory \ixo will have the necessary
  capabilities to deliver these measurements.
\end{abstract}

\maketitle

\section{Introduction}
One of the surprising legacies of the {\em Chandra} era is the
discovery of X-ray cavities in virtually every cool core galaxy
cluster \citep[e.g.][]{birzan:04,allen:06}.  This led to the
suggestion that AGN heating is responsible for halting the cooling in
clusters and keeping cooling flows from actually reaching star-forming
temperatures.

The discovery of the cavities allowed estimating the kinetic power of
a large sample of radio galaxies, to within about an order of
magnitude.  The remaining uncertainties result from the fact that the
cavity age cannot be determined from imaging studies alone.  A range
of age estimates have been used in the literature, ranging from the
sound crossing time to the buoyant rise time
\citep[e.g.][]{mcnamara:00,fabian:00,churazov:01}.  Additionally, the
orientation of the cavities relative to the line-of-sight cannot be
determined from imaging observations alone, leading to an additional
uncertainty in cavity size and age \citep[e.g.][]{ensslin:02c}.
Finally, given that we do not know the inflation history of cavities,
it is not clear how much energy has to be expended by the black hole
to inflate a cavity to a given size.

Thus, while it is now generally accepted that AGN can release enough
energy in principle to heat clusters (at least to within a factor of
order unity), it is not clear whether and how this energy is
transferred to the thermal gas in order to actually counteract the
cooling.

To make progress, a quantitative spectroscopic approach is necessary.
Unfortunately, neither {\em Chandra} nor {\em XMM-Newton} have the
high spectral resolution and throughput required to observe the
physical processes that control feedback.  As we will show below, the
requirements to solve the outstanding problems in cluster feedback
posed by {\em Chandra} will be met by the {\rm International X-ray
  Observatory, IXO}. Before presenting our simulations, it is worth
reviewing the relevant technical specifications currently planned for
{\em IXO}:
\begin{itemize}
\item{{\bf Spectral resolution:} {\em IXO} will feature an X-ray
  Micro-calorimeter instrument ({\em XMS}) with uniform 2.5eV spectral
  resolution up to 7 keV.  At the Fe K$\alpha$ line energy of $\sim
  6.7$ keV, this corresponds to a spectral resolving power of 2700, an
  improvement of roughly two orders of magnitude compared to the {\em
    Chandra} resolution for extended sources.  The core {\em XMS}
  array will have 3 arc-second pixels, arranged in a 40x40 square for
  a 2x2 arc-minute field-of-view.}
\item{{\bf Effective area:} Relative to {\rm Chandra} imaging with the
  back-illuminated S3 chip, the effective area will be roughly a
  factor 25 higher, corresponding to the efficiency difference between
  the Keck telescopes and the Hubble Space Telescope {\em HST}.}
\item{{\bf Angular resolution:} The angular resolution of the {\em
    IXO} telescope mirror assembly will be 5 arc-second on-axis at 7
  keV (5 arc-second half-power diameter), which is very well suited
  for cluster studies.  Again this is comparable to the angular
  resolution achievable by ground-based large telescopes compared to
  {\em HST}\footnote{While the 0.5 abscond {\em Chandra} resolution is
    necessary for the study of very fine details like the sound waves
    in Perseus, a fair fraction of cluster science enabled by Chandra
    could be done at 5 arc-second resolution.}.}
\end{itemize}

The organization of this letter is as follows: In \S\ref{sec:sims} we
describe the methods employed in the simulations presented in
\S\ref{sec:res}, while \S\ref{sec:dis} discussed the consequences and
\S\ref{sec:con} presents our conclusions. Throughout this letter, we
use concordance cosmological parameters of $H_{0}=70\,{\rm
  km\,s^{-1}\,Mpc^{-3}}$, $\Omega_{\rm m}=0.3$, and
$\Omega_{\Lambda}=0.7$ and cluster metallicities of 0.5 relative to
solar.

\section{Simulations}
\label{sec:sims}
A powerful way to make detailed observational predictions of feedback
is by constructing dedicated numerical simulations of radio galaxies
in clusters and then virtually observing them with an X-ray telescope.
This is the avenue chosen in this letter. Before presenting our
results, we will briefly describe the method used.

\subsection{The code}
We use the publicly available {\tt FLASH} code \citep{fryxell:00}
which is a modular block-structured adaptive mesh refinement code.  It
solves the Riemann problem on a Cartesian grid using the
Piecewise-Parabolic Method. Our simulation includes $7\times 10^5$
dark matter particles. The particles are advanced using a cosmological
variable-timestep leapfrog-method.  Gravity is computed by solving
Poisson's equation with a multigrid method using isolated boundary
conditions. Our simulations have an effective dynamic range between
10000 and 20000.

In order to simulate radio galaxies in clusters, we inject
back-to-back supersonic jets by placing arbitrary inflow-boundary
conditions into the grid.  The location of the ``jet nozzle'' tracks
the dynamical center of the cluster potential.  The jets are injected
with a velocity of 0.1c and an internal Mach number of 37.

In order to reproduce the observed morphologies we impose a sub-grid
jitter on the jet, confined to a 20 degree opening angle, known as the
dentist drill effect \citep{scheuer:82}.  More information on the
numerical details of the type of simulations used in this letter are
presented in \cite{heinz:06b}.

\subsection{Initial conditions}
We used the S2 galaxy cluster from \cite{springel:01} as our initial
conditions.  This is a cosmologically evolved rich cluster with a mass
of $M\sim 10^{15}\,{M_{\odot}}$.  We imported the initial conditions
into FLASH and evolved the cluster for several dynamical times to
eliminate any transients before setting off the jet.

The density of the S2 cluster is appropriate for all the clusters
presented in this paper.  The temperature profile is appropriate for
our simulations of Cygnus A and Hydra A, but we had to adjust the
temperature normalization for our simulations of the Perseus cluster,
which has a moderately lower temperature.

Because the cluster is fully cosmologically evolved, our simulations
incorporate a realistic level of density and temperature
sub-structure, as well as a realistic cluster velocity field,
including turbulence.  We did not, however, include a sub-grid
turbulence model.

\subsection{{\tt XIM:} a virtual observatory}
The simulation output was then virtually observed using the publicly
available in-house tool \xim \citep[see][]{heinz:09}.  Taking input
grids from numerical hydrodynamic simulations, \xim performs spectral
modeling of thermal emission, including Doppler shifts and ionization
balance.  It then performs spectral projection along an arbitrary
line--of--sight, PSF convolution, telescope and detector efficiency,
and spectral convolution with the detector response (using the proper
response files for current and future telescopes). Finally, it adds
sky- and instrument backgrounds and Poisson counting error.

\subsection{Fe XXV as a kinematic tracer}
In this paper we concentrate on the K$\alpha$ line from Helium-like
iron Fe XXV.  This line complex is very luminous in the intra-cluster
gas and because of the large mass of Fe relative to other abundant
species, it is an excellent kinematic tracer \citep{brueggen:05}.

\begin{figure}[t]
\begin{center}
\resizebox{0.9\columnwidth}{!}{\includegraphics{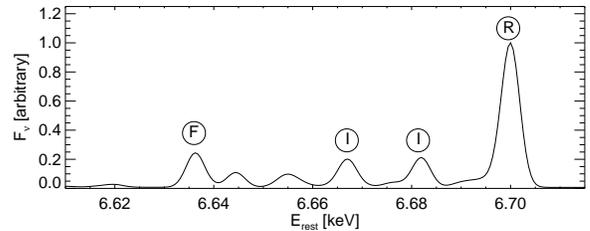}}
\end{center}
\caption{Reference spectrum of the Fe XXV K$\alpha$ line for a 4keV
  thermal plasma, with resonance, intercombination, and forbidden
  lines labeled. Di-electronic satellite lines are also
  included.\label{fig:fe}}
\end{figure}

Many other lines are present, also showing the kinematic signatures
discussed below, but for simplicity and clarity we limited the
analysis to Fe XXV.  Figure~\ref{fig:fe} shows the rest-frame
multiplet line structure at a plasma temperature of 4 keV for
reference.  Given its strength, the resonance line at 6.7 keV rest
energy is the best suited for kinematic measurements.

\section{Results}
\label{sec:res}
To investigate the possibilities offered by high-throughput X-ray
spectrographs for the study of AGN feedback in clusters, we
constructed virtual \ixo observations of three representative
(benchmark) galaxy clusters with detailed {\em Chandra} cavity
studies: Perseus A, Cygnus A, and Hydra A.  The simulations aim to
reconstruct, as best as currently possible, the observational
characteristic of these sources.

Before presenting the three cases in more detail, it is worth pointing
out that \ixo data are difficult to visualize in 2D figures.  The plots
show different spectra from virtual spectral slits placed across that
data cube, which clearly show the ability of \ixo to resolve the
kinematic signatures of cluster feedback.  However, a much better way
to view the data is by animation.  For this reason we have prepared
movies of all three cases that can be viewed at
http://www.astro.wisc.edu/$\sim$heinzs/feedbackmovies.html.

\subsection{Perseus A}
The Perseus cluster was the first cluster known to have X-ray
cavities.  With a mega-second of {\em Chandra} observing time, this is
the best studied cluster to date, displaying rich morphological
complexity across all wavebands
\citep[e.g.][]{graham:08,fabian:08,taylor:06}.  As such, it represents
the benchmark for numerical models.

As a moderately powerful source, with a jet power somewhere between
$10^{44}$ and $10^{45}\,{\rm ergs\,s^{-1}}$, it also represents a good
example of what has become known as ``gentle'' or ``effervescent''
heating \citep[e.g.][]{ruszkowski:04}.

The detection of a weak shock in the long {\rm Chandra} image
\citep{graham:08} has provided the most robust power estimate of
$W_{\rm jet}\sim 10^{45}\,{\rm ergs\,s^{-1}}$, which is the number we
chose to use.  Given this power estimate, we ran the simulation for
$10^7\,{\rm yrs}$ until the size of the shock and cavities
corresponded to the observed values.

The central cluster temperature is around 3-4 keV, somewhat colder
than the S2 cluster we used as our initial condition.  We thus
adjusted the cluster temperature in post-processing to ensure the
correct line strengths and ratios, fixing the total X-ray power from
the cluster to the observed {\rm Chandra} flux.

\begin{figure}[t]
\begin{center}
\resizebox{0.9\columnwidth}{!}{\includegraphics{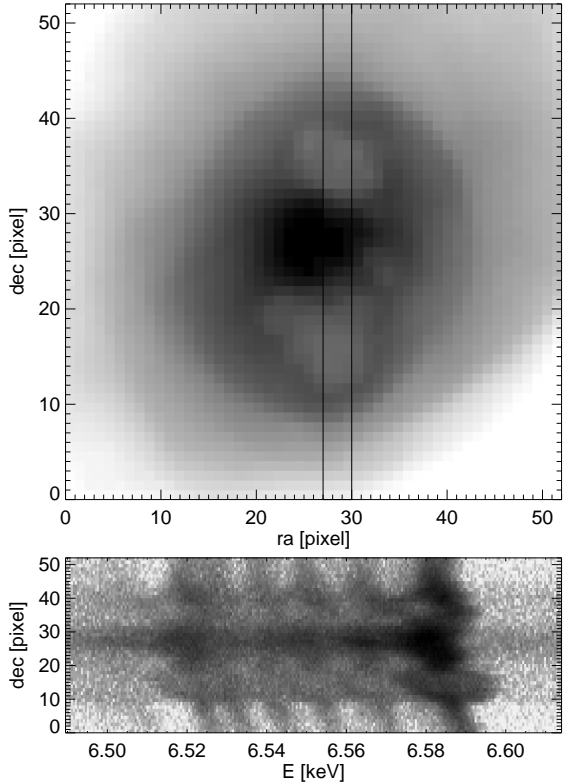}}
\end{center}
\caption{Virtual 250 ksec \ixo observation of Perseus cluster
  (redshift $z=0.01756$).  Top: 0.5-10 keV image. Bottom: Fe XXV
  K$\alpha$ spectrum of a 2-pixel vertical slit across the
  cavities, clearly showing the kinematic signature of the expanding
  bubble.\label{fig:perseus}}
\end{figure}

The resulting simulated \ixo image and Fe XXV K$\alpha$ spectrum are
shown in Fig.~\ref{fig:perseus} for an assumed exposure time of 250
ksec, at a red-shift of $z=0.01756$.  The cavities are clearly
visible, as is the classical bright central core of the Perseus
cluster.

Figure \ref{fig:perseus} also show the spectrum for a virtual spectral
slit placed across both cavities (see image for slit placement).
Apart from the complex sub-structure of the Fe XXV K$\alpha$
multiplet, the most striking feature immediately visible from the
spectrum is how the lines splits at the locations of the cavities.

The two ``bubbles'' seen in the spectrum correspond to the front and
back walls of the cavities, allowing full 3D reconstruction from the
imaging spectrum like in the case of supernova remnants.  Simply
reading off the expansion velocity from the blue- and red-shifted line
gives a line-of-sight velocity of $v_{\rm LOS} \sim \pm 350\,{\rm
  km\,s^{-1}}$, compared to the actual velocity of $\sim 325\,{\rm
  km\,s^{-1}}$ in the simulation.

Thus, \ixo will be able to clearly resolve the kinematic signature of
expanding bubbles around cluster radio sources like Perseus A.  It is
also clear that angular resolution of a few arc-seconds is vital to
resolve the cavities both spectrally and spatially.

\subsection{Cygnus A}
As a powerful FR2 source, Cygnus A sits at the upper end of the
expected range of cluster radio sources and shows clear cavities and
possibly a shock in the {\em Chandra} image \citep{wilson:06}.  As
with most cluster sources, the ambiguities in the interpretation of
the imaging data cannot constrain the jet power to better than about
an order of magnitude, roughly $10^{45}$ to $10^{46}\,{\rm
  ergs\,s^{-1}}$.

The discovery of giant X-ray cavities in other clusters has shown that
powerful outbursts like the one currently observed in Cygnus A might
be more common than implied by models of ``gentle'' cluster heating,
making this source an important benchmark.

Our initial numerical simulations of Cygnus A have been presented in
\cite{heinz:06b}.  Here we will present the virtual \ixo observation
derived from that simulation.  We ran the simulation with a jet power
of $10^{46}\,{\rm ergs\,s^{-1}}$ for 21 Myrs, at which point the
simulation reached the observed cavity and radio lobe size and
morphology.\footnote{We also tested the predicted virtual {\em
    Chandra} surface brightness of the simulation against the actual
  observation and found that they agree to within better than 10\% in
  the region of the rim around the cavity, giving us confidence in the
  predicted line fluxes.}

\begin{figure}[t]
\begin{center}
\resizebox{0.9\columnwidth}{!}{\includegraphics{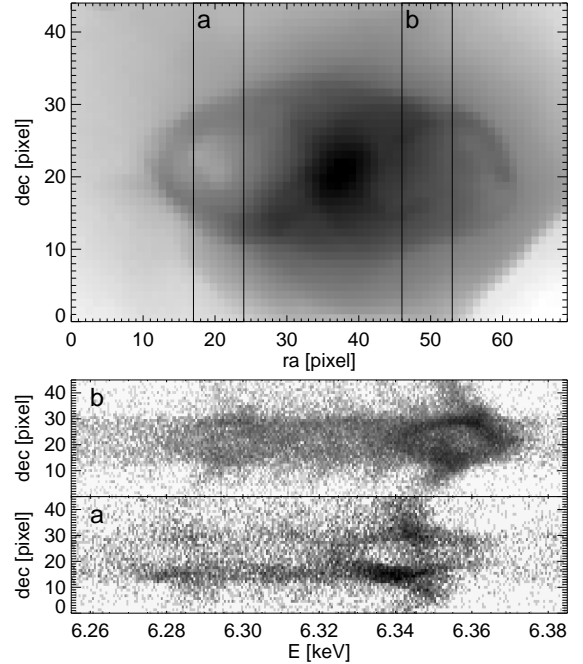}}
\end{center}
\caption{Virtual 250 ksec \ixo observation of Cygnus A (redshift
  $z=0.056$). Top: 0.5-10 keV image. Middle: Fe XXV K$\alpha$ line of
  the eastern cavity. Bottom: same for the western cavity.  Both
  cavities will be easily resolved by \ixo.\label{fig:cygnus}}
\end{figure}

Our virtual 250 ksec \ixo observation is shown in
Fig.~\ref{fig:cygnus}, for a cluster redshift of $z=0.056$.  The two
bottom spectra show the FeXXV K-alpha line, clearly resolving the two
cavities (the two spectra correspond to two virtual spectral slits,
one for each cavity).  As expected, for a source this powerful it will
be easy for \ixo to resolve the kinematic structure of the line.
Thus, \ixo will be able to conclusively determine the jet power of
Cygnus A.

\subsection{Hydra A}
Several clusters that currently contain relatively modest radio
sources show fossil cavities that imply much more powerful past
outbursts, most notably Hydra A \citep{wise:07}, MS 0735
\citep{gitti:07}, and Hercules A \citep{nulsen:05}.  Confirming this
interpretation will be an important test of impulsive vs.~gentle
cluster heating models (note that \cite{wise:07} identify multiple
powerful outbursts in Hydra A).

We used a late timestep of our $10^{46}\,{\rm ergs\,s^{-1}}$
simulation to represent this case, 150 Myrs after the outburst
started, roughly corresponding to the case of Hydra A.  The cavity
sizes are of the same order as those observed in Hydra A, including
the north-south asymmetry due to motion of the intra-cluster medium
(ICM).

\begin{figure}[t]
\begin{center}
\resizebox{0.9\columnwidth}{!}{\includegraphics{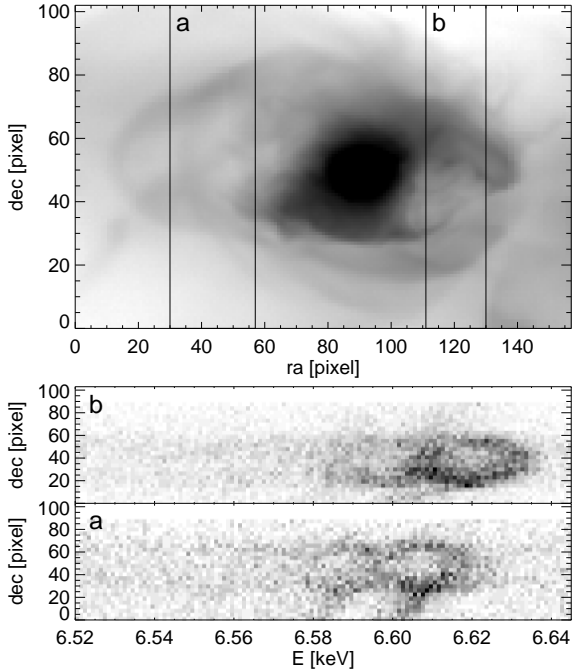}}
\end{center}
\caption{Virtual 250 ksec \ixo observation of a Hydra A--like radio
  source.  Top: 0.5-10 keV image. Middle and bottom: Fe XXV K$\alpha$
  spectra of both cavities.\label{fig:hydra}}
\end{figure}

Given the low surface brightness this far out in the cluster and the
large angular size of these sources, such observations will be more
challenging (requiring multiple long pointings).  Still, the 250 ksec
virtual observation in Fig.~\ref{fig:hydra} shows that \ixo can
resolve and detect the velocity structure of the cavities in these
sources.

\section{Discussion}
\label{sec:dis}
\subsection{Implications for other cluster radio sources}
The three detailed examples presented above show that a \ixo will be
able to resolve the kinematic structure of radio galaxy drive cavities
for a significant fraction of the sources.  In the case of Perseus A,
the large count rate will allow a clear resolution of the iron line in
observations as short as 20 ksec. This implies that even in clusters
with lower surface brightness, \ixo would be able to deliver cavity
velocities in exposures of moderate length (100-200 ksec).

In principle, \ixo will be able to resolve cavities expanding with
velocities as slow as $v_{\rm LOS} \gtrsim 100\,{\rm km\,s^{-1}}$,
though the presence of turbulent broadening in the ICM will likely
mean a higher threshold (see \S\ref{sec:caveats}).  Given that the
expected expansion velocity of actively jet-driven cavities for a
fixed physical cavity radius $R$ scales like \citep{heinz:98}
\begin{equation}
  v_{\rm LOS} \propto \left(\frac{W_{\rm jet}}{\rho_{\rm
      ICM}}\right)^{1/3}
\end{equation}
low--power sources like M87 would be more difficult to observe.
Detailed predictions of what \ixo would observe in the case of M87
will have to await more dedicated simulations.

\subsection{Caveats}
\label{sec:caveats}
Given that the presented virtual observations are based on numerical
simulations, there is are several physical effects that we did not
include and that could in principal affect the predicted results.
However, the point of this paper is not the precise prediction of
velocities of individual sources but rather to demonstrate that \ixo
will be able to measure the expansion rates for the expected range of
velocities.  We will therefore not go into any detail of the numerical
shortcomings (like the absence of magnetic fields, the limitation to
non-relativistic simulations, the restriction to purely thermal
particle populations, and neglecting effects of radiative transfer
like resonant scattering).

One of the critical elements that should be addressed is the level of
{\em turbulence} present in the ICM significant turbulence will
broaden all lines and make kinematic measurements of the sort
discussed here more difficult \citep{inogamov:03,churazov:04}.  While
our simulations do include turbulence on the scales resolved by our
grid, we cannot make any statements regarding the level of sub-grid
turbulence.  Turbulent velocities in excess of $300\,{\rm km\,s^{-1}}$
will impede cavity measurements. 

Finally, calibration uncertainties present in any X-ray instrument
will in principle affect the spectroscopic accuracy.  However, because
we are only discussing kinematic effects, systematic errors due to
calibration errors should not have any strong effect on our results.

\subsection{Other prospects for feedback studies}
The analysis presented above is only the most forward and simple
application of high-resolution, high-throughput X-ray spectroscopy to
the questions of cluster feedback.  Given the spectral complexity of
typical thermal spectra, many more applications will be possible that
are well beyond the scope of this letter.  We will suggest a few
avenues that promise to be particularly fruitful:
\begin{itemize}
\item{{\bf Line stacking:} Given \ixo's high spectral resolution over
  a wide bandwidth, it will be easy to re-grid the spectral data to a
  logarithmic grid, upon which one can stack different emission lines
  on top of each other simply by shifting along the energy axis,
  greatly increasing signal-to-noise (thus reducing the overall
  exposure time requirements).  More complex deconvolution algorithms
  could yield even more powerful diagnostics, including the
  temperature structure of the cavity shells from line-ratio
  variations.}
\item{{\bf Line-of-sight angles:} From the velocity centroid of the
  two cavities, relative to the cluster mean, it will be possible to
  determine the mean jet orientation relative to the line-of-sight as
  well as detecting large scale rotation in the cluster.}
\item{{\bf Emission measure mapping:} Beyond simple kinematics,
  spectral fitting will allow the extraction of the emission measure
  distribution in each individual X-ray pixel, allowing multi-phase
  metallicity, entropy, temperature, and pressure maps.  The ultimate
  promise of this approach is the direct detection of the
  spectroscopic signature of AGN heating.  The performance
  requirements on the {\em XMS} calibration to allow such a
  measurement has yet to be formulated.}
\end{itemize}
 
\section{Conclusions}
\label{sec:con}
We showed that the quantitative kinematic measurements critical for
establishing accurate cavity ages, jet powers, and black hole duty
cycles are only achievable through high-resolution X-ray spectroscopy.
Our simulations demonstrate that the 2.5 eV spectral resolution and 5
arc-second spatial resolution provided by the International X-ray
Observatory \ixo will be sufficient to make the necessary
measurements.  Any significant compromise in angular or spectral
resolution would render these measurements impossible, implying that
neither the current or any other planned X-ray telescopes (such as
{\em Astro-H}) could deliver them instead of \ixo.

\end{document}